\documentclass[manuscript]{aastex}
\usepackage{graphicx}

\shorttitle{Infrared Imaging of Capella with the IOTA}
\shortauthors{Kraus et al.}
\received{2004-09-10}
\accepted{2005-03-21}

\begin{document}

\title{Infrared Imaging of Capella\\with the IOTA Closure Phase Interferometer}

\author{S.~Kraus\altaffilmark{1, 2}, F.~P.~Schloerb\altaffilmark{1},
  W.~A.~Traub\altaffilmark{3}, N.~P.~Carleton\altaffilmark{3},
  M.~Lacasse\altaffilmark{3}, M.~Pearlman\altaffilmark{3},
  J.~D.~Monnier\altaffilmark{4}, R.~Millan-Gabet\altaffilmark{5},
  J.-P.~Berger\altaffilmark{6}, P. Haguenauer\altaffilmark{7},
  K. Perraut\altaffilmark{6}, P. Kern\altaffilmark{6}, 
  F. Malbet\altaffilmark{6}, P. Labeye\altaffilmark{8}}
\email{skraus@mpifr-bonn.mpg.de}

\altaffiltext{1}{Department of Astronomy, University of Massachusetts, Amherst
  MA 01003, USA}
\altaffiltext{2}{Max-Planck Institut f\"ur Radioastronomie (MPIfR), Auf dem
  H\"ugel 69, 53121 Bonn, Germany}
\altaffiltext{3}{Harvard-Smithsonian Center for Astrophysics (CfA), 60 Garden
  Street, Cambridge, MA 02138, USA}
\altaffiltext{4}{Astronomy Department, University of Michigan, 500 Church
  Street, Ann Arbor, MI 48104, USA}
\altaffiltext{5}{Michelson Science Center (MSC), California Institute of Technology, Pasadena, CA 91125, USA}
\altaffiltext{6}{Laboratoire d'Astrophysique de Grenoble (LAOG), 414 Rue de la Piscine, BP 53X, 38041 Grenoble Cedex, France}
\altaffiltext{7}{Jet Propulsion Laboratory (JPL), California Insititute of Technology, MS 306-388, 4800 Oak Grove Drive, Pasadena, CA 91109, USA}
\altaffiltext{8}{LETI-CEA D\'{e}partement de Microtechnologies, 17 rue des Martyrs, 38054 Grenoble Cedex 9, France}

\newcommand{\magn}[2]{\ensuremath{#1 \overset{m}{.} #2}}

\frenchspacing

\begin{abstract}
We present infrared aperture synthesis maps produced with the upgraded IOTA
interferometer. Michelson interferograms on
the close binary system Capella ($\alpha$ Aur) were obtained in the $H$-band
between 2002 November 12 and 16 using the IONIC3 beam combiner.
With baselines of $15$\,m $\le B \le 38$\,m,
we were able to determine the relative position of the binary components with
milliarcsecond (mas) precision and to track their movement along the $\approx14\degr$ arc
covered by our observation run. We briefly describe the algorithms used for
visibility and closure phase estimation. Three different Hybrid Mapping and
Bispectrum Fitting techniques were implemented within one software framework and used
to reconstruct the source brightness distribution. 
By dividing our data into subsets, the system could be mapped at three
epochs, revealing the motion of the stars. The precise position of the
binary components was also determined with model fits, which in addition
revealed $I_{Aa}/I_{Ab}=1.49 \pm 0.10$ and apparent stellar uniform-disk (UD)
diameters of $\Theta_{Aa}=8.9 \pm 0.6$ mas and $\Theta_{Ab}=5.8 \pm 0.8$ mas.

To improve the $u, v$-plane coverage, we compensated this orbital motion by
applying a rotation-compensating coordinate transformation. The resulting
model-independent map with a beam size of $(5.4 \times 2.6)$ mas allows the
resolution of the stellar surfaces of the Capella giants themselves.
\end{abstract}


\section{Introduction}

Aperture synthesis images, obtained using an interferometer with multiple
baselines and closure phases, provide a much more assumption-free way to
interpret interferometric data astrophysically than the simple fitting of
models to single-baseline fringe amplitudes does.
Therefore, imaging capabilities will be of special importance for the
upcoming generation of ground-based and space interferometers. In this paper,
we present the 
results of an initial attempt to use the \textit{Infrared Optical Telescope
Array} (IOTA) for imaging in the near infrared. Since this was to be the first
effort to construct an image at IOTA, we decided that it was important to
observe an object with a well known brightness structure in order to
demonstrate the instrument's capabilities. Therefore, we chose to observe the
strongly resolved spectroscopic binary Capella ($\alpha$ Aurigae) so that our
maps could be compared to the results of earlier studies.

The first spectroscopic measurements by~\cite{cam1899} revealed the components
Capella Aa and Ab (classified today as G8 III and G1 III \citep{joh02}). The
system also became the first binary to be separated by an optical
interferometer \citep{and20}. The orbit was determined with high precision in
1994 using the Mark III interferometer by~\cite{hum94}. In 1996 Baldwin et
al. presented the first optical ($\lambda=830$ nm) aperture synthesis map of
Capella using the \textit{Cambridge Optical Aperture Synthesis Telescope}
(COAST) with a longest baseline of $B=6.1$m. Two years later Young used the
same instrument to map Capella in the near infrared \citep[$\lambda=1.3$
$\mu$m;][]{you99}.\\

We observed Capella on five nights in November 2002 in the $H$-band
($\lambda=1.65 \mu$m; $\Delta\lambda=0.3 \mu$m). In this paper, we review the
data reduction procedure used to determine visibilities and closure phases
from the raw interferometer data, and we present model fits and images derived
from the visibility and closure phase data. By investigating the well-studied
Capella system, we have been able to calibrate the closure phase sign and to
determine the overall behaviour of this recently upgraded 3-telescope
interferometer. One of the mapping procedures described within this paper has
also been used to generate the maps presented in~\cite{mon04} depicting the
binary $\lambda$ Virginis.

\section{Observations}

\begin{deluxetable}{cccc}
\tabletypesize{\scriptsize}
\tablecaption{Observation Log\label{tab:obslog}}
\tablewidth{0pt}
\tablehead{
\colhead{Date [UT]} & \colhead{Start time [UT]} & \colhead{No. exposures\tablenotemark{1}} & \colhead{Calibrators\tablenotemark{2}}}
\startdata
  11/12/02 & 07:43 & 20 & $\alpha$ Cas, $\beta$ Aur \\
           & 10:11 & 27 & $\beta$ Aur \\
           & 12:00 & 16 & $\beta$ Aur \\
  11/13/02 & 07:19 & 26 & $\delta$ Aur, $\beta$ Aur \\
           & 09:09 & 20 & $\beta$ Aur, $\delta$ Aur \\
           & 10:45 & 24 & $\delta$ Aur, $\beta$ Aur \\
  11/14/02 & 05:51 & 25 & $\kappa$ Per, $\beta$ Aur \\
           & 08:24 & 21 & $\delta$ Aur, $\beta$ Aur \\
           & 10:25 & 39 & $\delta$ Aur, $\beta$ Aur \\
           & 12:06 & 19 & $\beta$ Aur, $\delta$ Aur \\
  11/15/02 & 07:08 & 25 & $\alpha$ Cas, $\delta$ Aur \\
           & 09:28 & 53 & $\delta$ Aur \\
           & 11:34 & 18 & $\delta$ Aur \\
           & 12:50 & 6 & $\delta$ Aur \\
  11/16/02 & 09:12 & 45 & $\delta$ Aur, $\beta$ Aur \\
           & 12:22 & 17 & $\delta$ Aur \\
\enddata
\\[2mm]
Interferometer Configuration used:\\
2002\,Nov\,12-14: A35, B15, C0\\
2002\,Nov\,15-16: A15, B15, C0\\ 
where A, B, C denotes the individual telescopes and the number of their
position\\ on the north-east (for A and C) or south-east baseline (for B) in
meters.
\tablenotetext{1}{Each exposure consists of 500 individual scans.}
\tablenotetext{2}{Compared to other calibrator stars, $\alpha$ Cas is clearly
  resolved. In order to still use it as a calibrator we applied a UD
  correction based on the reference diameter $\Theta_{\alpha
    \textnormal{Cas}}=6.25$ mas given by~\cite{ric02}. Also the well-known
  binary $\beta$ Aur with a semi-major axis of 3.3~mas appears marginally
  resolved to our interferometer. We used the orbital elements and UD
  diameters derived from measurements with the Mark III
  interferometer~\citep{hum95} to apply a correction to our measured
  visibilities and closure phases before calculating the transfer
  function. The short orbital period (3.96~d) requires to compute this
  correction for each of our measurements on $\beta$ Aur separately.} 
\end{deluxetable}

The IOTA was constructed jointly by the Smithsonian Astrophysical Observatory
(SAO), Harvard University, the University of Massachusetts (UMass), the
University of Wyoming, and MIT/Lincoln Laboratory \citep{tra02}. This
3-telescope interferometer located at Fred Whipple Observatory atop Mount
Hopkins, Arizona, demonstrated first closure phases on 2002 February
25. IOTA's telescopes can be moved on an L-shaped track and are mounted on
stations $\approx5$\,m and $\approx7$\,m apart. With track lengths of 15\,m
(south-east direction) and 35\,m (north-east direction), baselines up to 38\,m
can be formed. The collecting optics
consist of f/2.5 45\,cm Cassegrain primary mirrors, which are fed by
siderostats. The atmospherically induced motion of the image is compensated by
tip-tilt servo systems mounted behind the telescopes. Passing various
mirrors and path-compensating delay lines, the beams are deflected into the
laboratory, where the infrared component of the beam is coupled onto the
IONIC3 pair-wise beam combiner \citep{ber03}. This combines the beams symmetrically with a ratio of 50:50. 

For each baseline the beam combination produces two complementary outputs
which are simultaneously acquired by a PICNIC-camera~\citep{ped04}. Using
piezo scanners installed in two optical path delays, the optical path delay (OPD) can be
modulated to scan temporally through the interference fringe pattern.

Each observation on the target object is completed by the acquisition of four
calibration files. One of these files measures the camera
background signal with the light from all telescopes shuttered out. For the
other three files, the light of two telescopes is shuttered out
alternately. This allows us to measure the instrument transmission 
by determining the coefficients of the transfer~matrix
\citep[$\kappa$~matrix,~][]{for97}.

The observations presented in this paper were obtained at the IOTA between UT
2002 November 12 and 16. During this time we made observations in two
three-telescope configurations in order to obtain a reasonable sampling of the
$u, v$ plane for the Capella image. A log of the observations is presented in
Table~\ref{tab:obslog}.

\section{Data Reduction}

\begin{figure*}[hbp]
  (see Figure attached)
  \caption{Representative corrected interferograms, all taken 2002Nov13 on the
  BC baseline. The selected scans represent typical conditions with low
  noise ($\alpha$ Cas), high noise ($\delta$ Aur), and
  interfering resonance ($\alpha$ Lyn). The wavelet spectral
  density power spectra shown in the lower panels are calculated from the
  corresponding raw interferograms. On the ordinate of the CWT plots, the 
  scale quantity is given which is equivalent to the period and inversely
  proportional to the frequency of the corresponding rescaled wavelet. The
  contours demonstrate how our departitioning algorithm removes regions not
  connected to the area with the highest peak and separates the fringe peak
  from the piston and resonances for the visibility estimation.
  } \label{fig:rawcwt}
\end{figure*}

IOTA produces raw fringes by scanning the optical path delay through the
position of the white light fringe in the instrument. The fringes which are
observed must then be converted into visibilities and closure phases
for imaging and model fitting. We have reduced the raw fringe data in a
sequence of steps.

An important first step is to locate the fringes within the raw data scans in
order to determine the optical path delay in the IOTA system. If fringes are
found, the next step is to calibrate photometric fluctuations due to
atmospheric seeing. By using the transfer matrix, the coupling of the stars
to the optical inputs of the beam combiner and the internal coupling of the
beams within the combiner~\citep{for97} are taken into accont as well.

In order to form a photometrically calibrated scan,
we make use of the fact that for the 3-beam pairwise combination scheme used
the photometric information can be obtained for each measurement using linear
combinations of the interferometric signals~\citep[similar to the procedure
used by~][]{mon01}.  This requires a matrix inversion, which might be sensitive
to numerical instabilities, especially in the case of low signal-to-noise
($S/N$). For this reason, we smooth the scan by convolving it with a Gaussian
before solving for the photometry for each individual pixel.
The photometric calibration is performed for the two complementary signals of
each baseline separately. Remaining residuals are removed by subtracting the
two complementary outputs for each baseline.


The upper panels of Figure~\ref{fig:rawcwt}
illustrate how the data might look for different conditions after these
basic data reduction steps.

Each observation at IOTA consists of measurements of hundreds of fringes
obtained over a time period of a few minutes.  Typically, we have estimated
visibilities and closure phases for each individual realization of a fringe
and then averaged the results for many fringes to obtain the final
result.

\subsection{Visibility Estimation} \label{cha:estvis}

The scans through the white light fringe follow a particular functional form.
Assuming a bandwidth filter with idealized rectangular transmission, the instrumental
response expected for a point source is given by the product of a
sinc-envelope, with width inversely proportional to the bandwidth of the
input signal, and a cosine function representing the interference fringes.
The amplitude of the fringe envelope reflects the \textit{raw visibility}.
However, this measurement must be calibrated with observations of unresolved
reference stars or stars of known diameter in order to remove instrumental and
atmospheric effects.
Typically, each observation of a target source is bracketed by observations of
unresolved stars to provide a calibration of the instrument. In this work,
all target observations are enclosed by at least one such calibration
measurement on each side of the target observation. The results for the
calibration stars are interpolated linearly to calibrate the raw visibilities  
for the interferometric efficiency of the system, yielding the
\textit{calibrated target visibility}.

During our observation run we noticed a minor technical problem which became a
dominating issue for data reduction: the scanning piezo mirrors showed a
mechanical resonance (see 3rd panel of Figure~\ref{fig:rawcwt}), resulting in
multiple additional peaks within the power 
spectrum of the acquired scans. The data reduction procedure most commonly
used by the interferometry community \citep[e.g.][]{bal96} estimates the
visibility by integrating over the fringe peak within interferogram power
spectra after averaging and background subtraction. To improve the visibility
estimation for our resonance-influenced data, we implemented two alternative
methods to estimate the visibility amplitude.\\

Since the interfering resonance only affects the internal structure of the
fringes, we implemented a Fringe Envelope Fitting (FEF) algorithm which
filters the peaks of the fringe packet.
The position of these local minima and maxima is then fitted to the analytic
sinc-envelope function with the envelope amplitude, the width, and the
position as fitting parameters. For low $S/N$ this algorithm
tends to overestimate the fringe amplitude. Analyzing the resulting bias by
simulating fringes with additive Gaussian noise, we found a bias on the fitted
fringe amplitude which depends exponentially on the noise. Based on these
simulations we were able to correct the fitted visibilities down to $S/N \approx 1$.

Motivated by~\cite{seg03}, we also implemented a data reduction procedure
based on the Continous Wavelet Transform (CWT). In multiscale analysis, the
time signal is decomposed into the two-dimensional time-frequency space by
measuring the response of a translated and scaled mother-wavelet function to
the measured signal. Compared to the Fourier transform (FT), which is based on
periodical sinusoidal waves, the use of a localized mother wavelet allows the
precise localization of the power within the signal~\citep{tor98}.
For the mother wavelet, we used the Morlet function, which is given by a
sinusoidal multiplied with a Gaussian and therefore similar to the analytic
fringe function as noted above. The absolute square of the complex wavelet
function defines the wavelet spectral density spectrum. 

From this density spectrum, we remove all features below a particular
significance level (e.g. $35\%$ scaled to the peak).  Resonance (see
Fig.~\ref{fig:rawcwt}) or an uncompensated change in piston
during the fringe scan might result in several non-connected
areas.  Therefore, we run a filling algorithm which removes all regions
which are not connected to the area containing the main signal.
The integral over the remaining region is used as an estimate of the fringe
power.

The measured width of the remaining region in the OPD and scale domain
is then compared to the expected width and used as a selection criteria to
reject scans with a high piston. Even for good scans, the CWT is contaminated
by additive noise. To determine the level of this noise, we integrate in the
CWT along the OPD axis towards both directions, starting $5 \mu$m from the
border of the fringe region. The extend of this noise region along the scale
axis is given by the extend of the fringe peak region.
This averaged background noise level is subtracted from the fringe
power. Noticing that the CWT is a non-linear transformation, we confirmed the
proper behavior of our algorithm by performing extensive simulations.

The visibilities obtained with the FEF and CWT algorithm are in good
agreement, even if the CWT algorithm seems to be superior for cases with low
$S/N$. The accuracy of the measured visibilities is limited primarily by
calibration. For error estimation we include, beside the statistical errors
for each data set, a calibration error for each night which was estimated by
the scattering of the calibrator visibilities during that particular
night. For good seeing conditions this calibration error $\Delta
V_{calib}^{2}$ is typically around $\pm 1\%$ (2002\,Nov\,12-15), while for
worse conditions it may rise to $\pm 2\%$ (2002\,Nov\,16).

\subsection{Closure Phase Estimation} \label{cha:estcp}

Since we are observing on the three baselines $AB, BC, AC$ simultaneously, the
fringe phases $\phi_{AB}$, $\phi_{BC}$, $\phi_{AC}$ can be measured.
Simultaneous measurement of three phases in turn allows a new quantity, the
closure phase $\Phi$ \citep[CP, ][]{jen58}, to be estimated. The CP has
enormous value since it is unaffected by atmospheric propagation effects, and
through measurements of CP and visibility amplitudes, it is
possible to create aperture synthesis images. In the following we briefly
describe our methods for estimating this essential interferometric
observable. 

The estimation of CP begins with the location of the fringes for
each baseline within the raw scans. The phase of each fringe must be
determined within the same temporal window within the scan in order to assure
that all phases are measured at the same time and within the coherence time of
the atmosphere. Thus, we locate the median delay for the centers of the three
fringe packets and determine the phase of each baseline within a short 
window of width $3 \mu$m centered on that OPD. The FT of the windowed fringe
results in a strong peak at the frequency of the fringe. In the IOTA
three-baseline system, the frequencies of these peaks for the three fringes 
obey a ``closure'' relationship such that $\nu_{AB}+\nu_{BC}-\nu_{AC}=0$, and
if this relationship is not obeyed by the data, then there is a good chance
that one or more of the peaks is spurious. For these reasons, we reject scans
which do not obey this frequency closure relation. For scans that do pass
this test, we may then use the phases of the FT's for the three fringes to
compute the CP: $\Phi = \phi_{AB} + \phi_{BC} - \phi_{AC}$. Unresolved
calibrators, for which the CP is known to be zero, are used to define the
instrumental CP-offset.

Using the above algorithm, we have estimated closure phases for all of the
observations obtained during our run. From the scattering of the calibrator CP
over the individual nights we include calibration errors $\Delta \Phi_{calib}$
between $\approx 0.2\degr$ and $1.1\degr$. Under good seeing conditions, the
IOTA/IONIC3 instrument is able to measure CP with high precision with
systematic errors of the order of $1.0\degr$ or less. As a conservative value
we assume $\Delta \Phi_{sys} = 2.0\degr$.

\subsection{Investigating the Effect of Bandwidth Smearing} \label{cha:bwsmearing}

After presenting the details of our procedures for visibility and CP
estimation, we want to investigate a potential influence of the
\textit{bandwidth smearing effect} on our data.

For extended sources, this effect will reduce the complex degree of coherence
at those points of the brightness distribution that are separated in
delay from zero optical path delay $\tau$ by amounts comparable to the width of
the fringe packet. As a consequence, the detailed instrumental response will be
more complicated than the point-source response. Thus, estimation of
visibilities and closure phases using the point source response as a template
could lead to some systematic errors.

For a rectangular spectral bandpass profile, the complex degree of coherence $c$
is given by
\begin{equation}
  c(\tau) = \frac{\sin \left( \pi \tau \Delta\lambda / \lambda^2
  \right)}{\pi \tau \Delta\lambda / \lambda^2}. \label{eqn:degcoh}
\end{equation}
To estimate the loss of coherence for a particular projected baseline
$B_{\bot}$, the \textit{coherent field of view} $\Delta\alpha =
\tau / B_{\bot}$ can be defined as the angular separation for
which the fringe of an off-axis point source will be centered right on the
first zero point of the fringe envelope of the on-axis point source. 
In this case, just half of the fringe power from both sources will overlap.
Using Equation~\ref{eqn:degcoh}, we obtain $\Delta\alpha = \lambda^2 /
(B_{\bot} \Delta\lambda) \approx 49$ mas for
our longest baseline used (AB on the nights of UT2002Nov12-14). 
The comparison with the separation of the Capella stars ($s \approx 46$ mas)
shows that we just reach the limit where bandwidth smearing might play a
role.

Since in the regime of marginal bandwidth smearing the contrast decreases with
distance from zero OPD in a well-defined way, a correction can be applied.
In the given case we correct the visibility $V_{UD}(\Theta)$ of a uniformly
bright disk of diameter $\Theta$. Assuming totally coherent combination,
$V_{UD}$ can be written as 
\begin{equation}
  V_{UD}(\Theta) = \frac{2 J_{1}(k \Theta B_{\bot})}{k \Theta B_{\bot}}, \label{eqn:VisUD}
\end{equation}
where $J_{1}$ denotes the Bessel function of first kind and first order.

To determine the delay-affected visibility $V_{UD}^{BW}$, we define
\begin{equation}
  V_{UD}^{BW}(\tau) = c(\tau) V_{UD}
\end{equation}
In the case of a binary star with two UDs of diameter $\Theta_{Aa}$ and
 $\Theta_{Ab}$ the resulting interferogram is given by the composition
of two fringes with different delays $\tau_{Aa}$ and $\tau_{Ab}$. Since the
delay at which the combined fringe is evaluated depends on the brightness
ratio of the individual fringes, it is appropriate to evaluate the two
correction factors relative to an effective delay $\tau_{\textnormal{eff}}$, given by the center of
intensity of the combined fringe packet:
\begin{eqnarray}
  \tau_{\textnormal{eff}} & = & \frac{I_{Aa} \tau_{Aa} + I_{Ab} \tau_{Ab}}{I_{Aa}+I_{Ab}}\\
  V_{UD, Aa}^{BW} & = & c(\tau_{\textnormal{eff}}-\tau_{Aa}) V_{UD}(\Theta_{Aa})\\
  V_{UD, Ab}^{BW} & = & c(\tau_{\textnormal{eff}}-\tau_{Ab}) V_{UD}(\Theta_{Ab}).
\end{eqnarray}

The fact, that our procedure for closure phase estimation does not measure the
phases over the whole fringe packet could be of advantage in the regime of
marginal bandwidth smearing. As described in chapter~\ref{cha:estcp}, we
measure the phases within a narrow temporal window around the center of the
composite fringe where the interferograms overlap properly.

\section{Model Fitting}

Making use of the expression for the visibility of a delay-affected UD, the 
complex visibility $\mathbf{V}$ of a close binary system with a position
vector $\vec{x}$ can be written as
\begin{equation}
  \mathbf{V}(\vec{x}) = \frac{I_{Aa} V_{UD, Aa}^{BW}
    + I_{Ab} V_{UD, Ab}^{BW} e^{-ik\vec{B} \cdot \vec{x}}}{I_{Aa}+I_{Ab}},\label{eqn:visbinary}
\end{equation}
with $\vec{B}=(u, v)$ for the baseline vector and the wave number $k=2\pi /
\lambda$. In strictly formal notation, an additional overall phase factor
appears on the right-hand side of Equation~\ref{eqn:visbinary}. Since this
phase factor merely moves the fringe pattern underneath the envelope it
neither effects the visibility nor the CP and can be neglected.

We performed least square model fits, equally weighting $N_{V}$ individual
visibility and $N_{\Phi}$ CP measurements with
\begin{eqnarray}
  \chi^{2} &=& \chi_{V^2}^{2} + \chi_{\Phi}^{2}\\
  \chi_{V^2}^{2} &=& \sum \limits_{i=1}^{N_{V}} \left( \frac{V_{i}^{2}-V_{model}^{2}}{\sigma_{V^2}} \right)^{2} \\
  \chi_{\Phi}^{2} &=& \sum \limits_{i=1}^{N_{\Phi}} \left( \frac{\Phi_{i}-\Phi_{model}}{\sigma_{\Phi}} \right)^{2},
\end{eqnarray}
with the model visibilities $V_{model}$ and closure phases $\Phi_{model}$
calculated from the complex visibility:
\begin{eqnarray}
  V_{model}^{2}(u, v) &=& \left| \mathbf{V}(u, v) \right|^{2}\\
  \Phi_{model}(u_{1},\ldots,v_{3}) &=& \sum_{i=1}^{3} \tan^{-1}
  \frac{\Im(\mathbf{V}(u_{i}, v_{i}))}{\Re(\mathbf{V}(u_{i}, v_{i}))}.
\end{eqnarray}
The errors for the individual measurements are $\sigma_{V^2}$ and $\sigma_{\Phi}$.
The model for Capella included the relative positions of the stars, the
stellar diameters, and the ratio of the total fluxes from the two stars.

\begin{figure*}[hbp]
  \center
  \includegraphics[width=450pt]{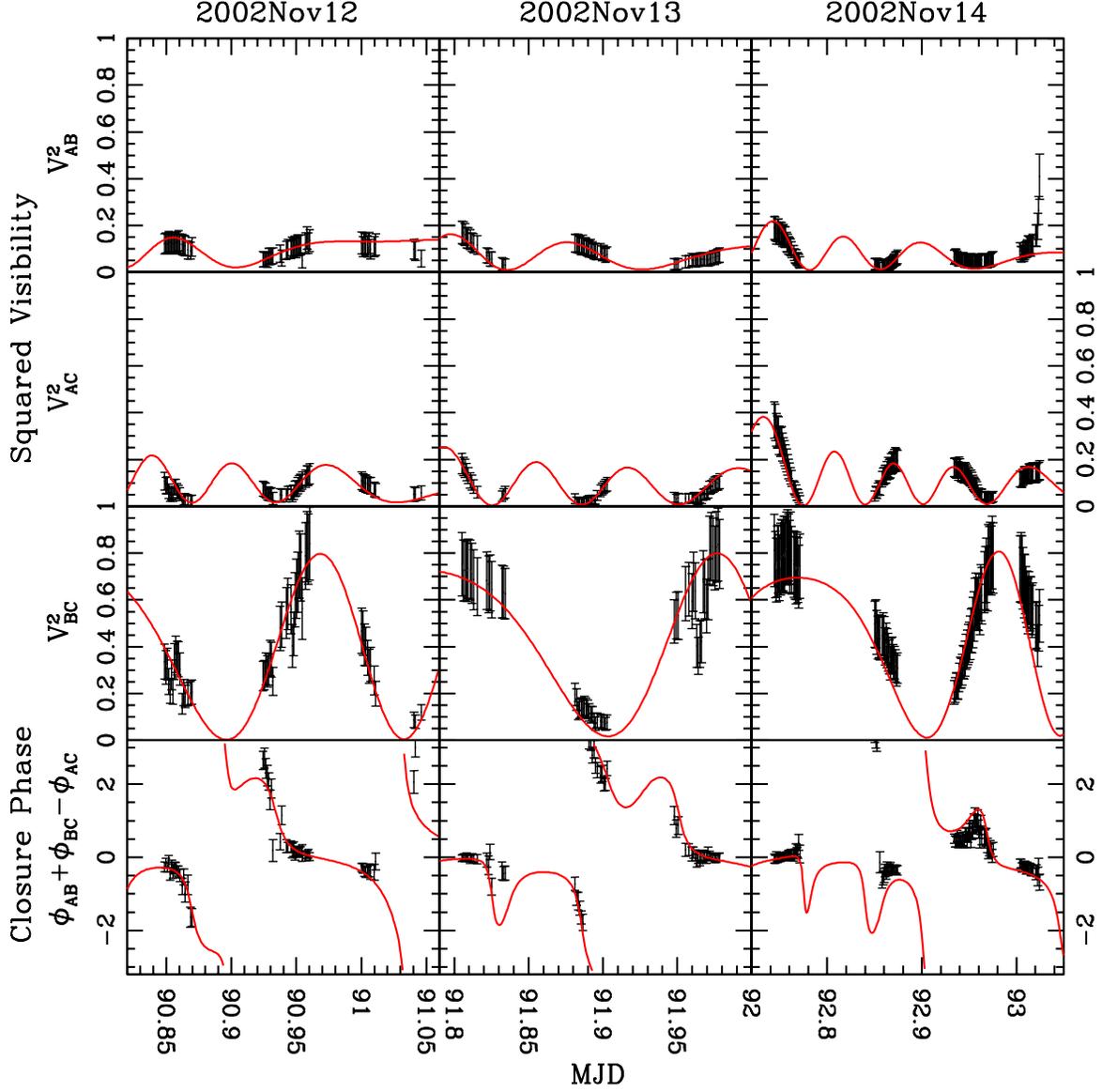}

  \caption{Best-fit models for Capella assuming uniform disks with parameters
  as listed in Table~\ref{tab:fitresults}. For all plots, the error bars shown
  include statistical, calibration, and systematic errors as discussed in
  chapters \ref{cha:estvis} and \ref{cha:estcp} and the CP is
  given in radians. The abscissa depicts MJD=JD$-$2452500. } \label{fig:modelfit}
\end{figure*}

Given our limited dataset, the Capella model contains parameters with
significant correlations, which makes it difficult to extract all parameters
simultaneously. Therefore, we proceeded to estimate the model parameters in an
iterative manner, starting without delay-compensated visibilities (using
$V_{UD}$ instead of $V_{UD}^{BW}$ in Equation~\ref{eqn:visbinary}). 
In the first step, the stars were assumed to be point
sources and the fit solved for the relative positions and fluxes. The next
step was to fit the residuals to this model fit and determine the stellar
diameters. The process continued iteratively, using the derived diameters to
find relative fluxes and positions followed by a new estimation of the
diameters. This process converged typically within a few iterations. Finally,
we applied the delay-compensation to obtain the fit results for each
individual night as given in Table~\ref{tab:fitresults} and
Fig. ~\ref{fig:modelfit}.

\begin{deluxetable}{ccccccccccccc}
  \rotate
  \tabletypesize{\scriptsize}
  \tablecolumns{12}
  \tablecaption{Fitting Results\label{tab:fitresults}}
  \tablewidth{0pt}
  \tablehead{
  \multicolumn{2}{c}{Data from Night [UT]} & \multicolumn{2}{c}{Data Points} &
  \multicolumn{2}{c}{$\chi^{2}$/DOF} & \colhead{Fitted\tablenotemark{1}} &
  \multicolumn{2}{c}{Fitted Diameter\tablenotemark{1}} &
  \multicolumn{2}{c}{Fitted Position\tablenotemark{2}} &
  \multicolumn{2}{c}{Ref. Pos.\tablenotemark{2, 3}}\\
  \colhead{Date} & \colhead{MJD=JD$-$2452500} & \colhead{$N_{V}$} &
  \colhead{$N_{\Phi}$} & \colhead{$\frac{\chi_{V}^{2}}{N_{V}}$} &
  \colhead{$\frac{\chi_{\Phi}^{2}}{N_{\Phi}}$} & $\frac{I_{Aa}}{I_{Ab}}$ &
  \colhead{$\Theta_{Aa}$ [mas]} & \colhead{$\Theta_{Ab}$ [mas]} & \colhead{dRA
  [mas]} & \colhead{dDEC [mas]} & \colhead{dRA [mas]} & \colhead{dDEC [mas]}}
  \startdata
  11/12/02 & 90.821 ... 91.046 & $165$ & $54$  & $0.76$ & $1.15$ & $1.24^{+0.36}_{-0.31}$ & $8.6^{+1.0}_{-0.8}$ & $5.9^{+1.3}_{-1.6}$ & $-9.28 \pm 1.94$  & $44.79 \pm 1.25$ & $-8.93$ & $44.70$ \\
  11/13/02 & 91.805 ... 91.978 & $173$ & $59$  & $1.11$ & $2.49$ & $1.51^{+0.29}_{-0.31}$ & $9.6^{+1.4}_{-1.2}$ & $5.2^{+3.4}_{-2.9}$ & $-12.14 \pm 1.68$ & $43.33 \pm 0.89$ & $-11.67$ & $43.25$ \\
  11/14/02 & 92.744 ... 93.026 & $315$ & $104$ & $1.05$ & $3.75$ & $1.42^{+0.25}_{-0.20}$ & $8.7^{+1.1}_{-1.0}$ & $6.0^{+2.0}_{-2.5}$ & $-14.29 \pm 1.24$ & $41.32 \pm 0.71$ & $-14.47$ & $41.60$ \\
  11/15/02 & 93.797 ... 94.042 & $363$ & $105$ & $0.29$ & $0.75$ & $1.50^{+0.20}_{-0.10}$ & \textemdash & \textemdash & $-17.90 \pm 3.90$ & $39.27 \pm 4.43$ & $-17.33$ & $39.72$ \\
  11/16/02 & 94.883 ... 95.034 & $183$ & $58$  & $0.56$ & $0.24$ & $1.68^{+0.29}_{-0.26}$ & \textemdash & \textemdash & $-20.24 \pm 1.40$ & $37.68 \pm 2.40$ & $-20.13$ & $37.68$ \\
  \enddata
  \tablenotetext{1}{To compensate the perceivable intra-night motion, the
  same rotation-compensating coordinate-transformation as described in
  chapter~\ref{cha:rotcomp} was applied separately to the data of each night
  before fitting the intensity ratio and the diameter.}
  \tablenotetext{2}{Relative positions are measured from the infrared brighter to the infrared fainter component.}
  \tablenotetext{3}{Reference orbit by~\cite{hum94}.}
\end{deluxetable}

\begin{figure*}[hbp]
  \center
  \includegraphics[width=450pt]{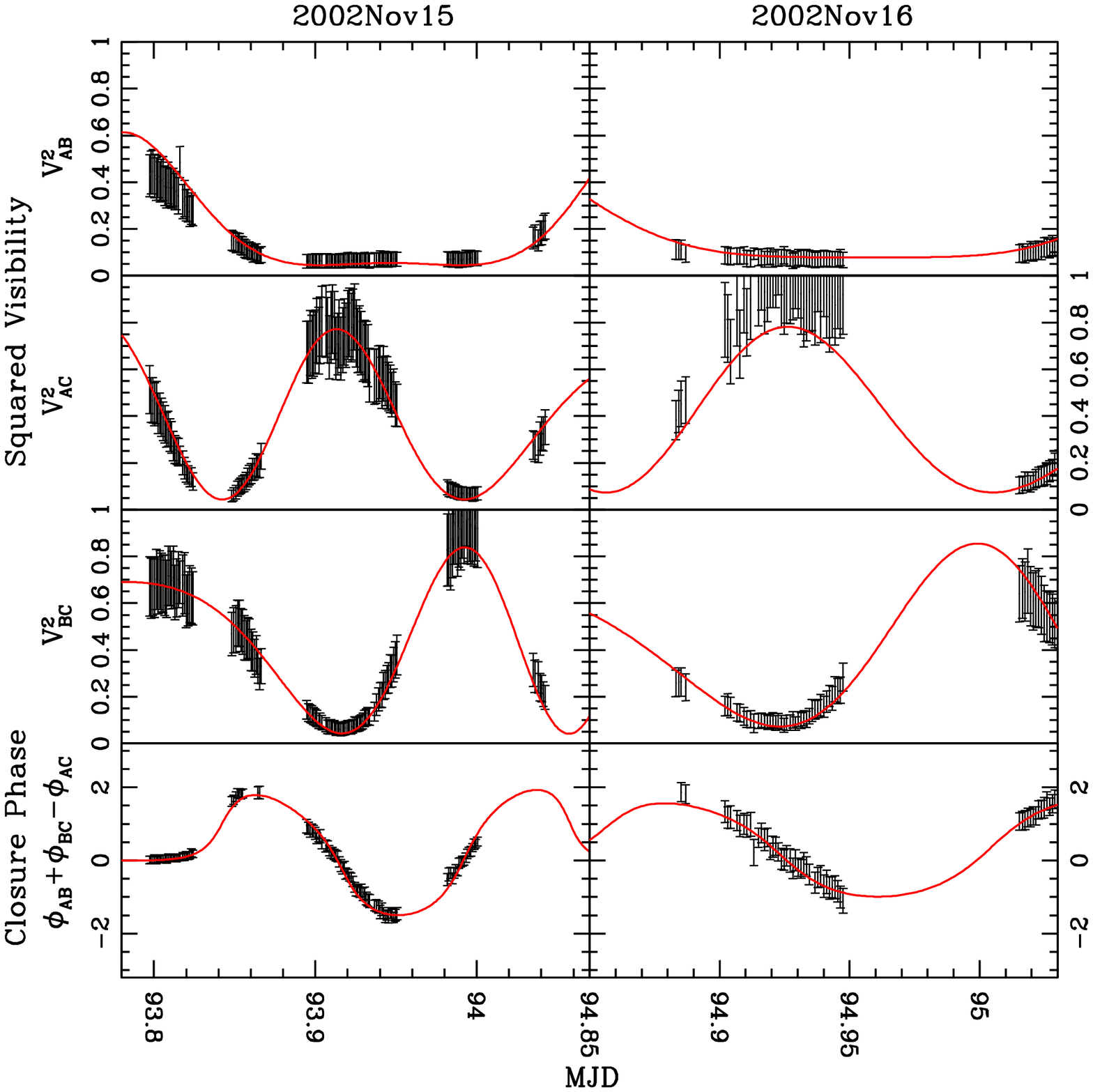}

  \caption{Best-fit models for Capella assuming uniform disks with parameters
  as listed in Table~\ref{tab:fitresults}. For all plots, the error bars shown
  include statistical, calibration, and systematic errors as discussed in
  chapters \ref{cha:estvis} and \ref{cha:estcp} and the CP is
  given in radians. The abscissa depicts MJD=JD$-$2452500. } \label{fig:modelfit2}
\end{figure*}

The fit results for each individual night are given in
Table~\ref{tab:fitresults} and Figures~\ref{fig:modelfit} and~\ref{fig:modelfit2}. In some cases, the
reduced $\chi^2$ values exceed unity, which might be a result of bad seeing,
a potentially uncorrected bias due to the effect of bandwidth smearing, the
piezo scanner resonances, or other undiscovered systematic errors. However,
the measurements are clearly in good agreement with the high precision
reference orbit by~\cite{hum94} obtained with two telescopes on the Mark III
interferometer.

Weighted averaging give for the diameters of the Capella giants,
$\Theta_{Aa}=8.9 \pm 0.6$\,mas and $\Theta_{Ab}=5.8 \pm 0.8$\,mas and the
intensity ratio $I_{Aa}/I_{Ab}=1.49 \pm 0.10$. Using the orbital parallax of
$75.0 \pm 0.57$\,mas \citep{pou00} we
calculate physical radii of $R_{Aa}=12.6 \pm 0.9$\,$R_{\Sun}$ and $R_{Ab}=8.3
\pm 1.1$\,$R_{\Sun}$. 

Stellar radii measurements are of special importance as the most direct way to
determine stellar effective temperatures. The stellar radii, derived from our
$H$-band data, are in good agreement with the measurements at visible
wavelengths by~\cite{hum94}. As a result of our longer effective wavelength
and the smaller amount of data, we cannot exceed the precision of their
extensive study. Therefore, we skip re-calculating the effective temperatures
with our values.

\section{Imaging}

\begin{figure*}[hp]
  \centering
  \epsscale{1.0}
  \plotone{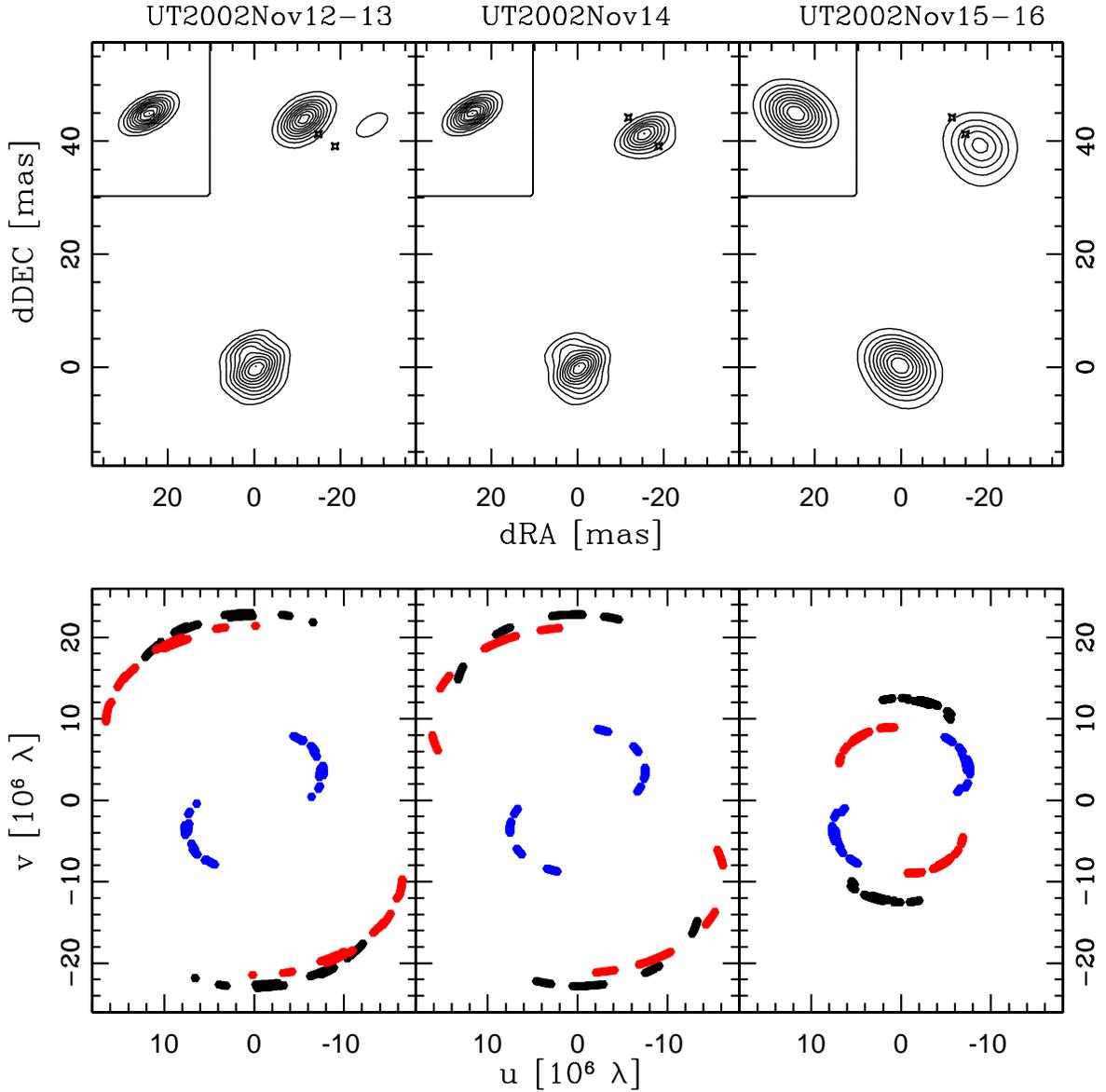}
  \caption{Image reconstructions from subsets of our complete data set revealing the orbital motion of the stars between three epochs. To compensate the marginal motion within the subsets, the $u, v$-plane was rotated synchronously to the reference position of the components at the intermediary modified julian dates MJD=$91.437$, $93.907$ and $94.262$ (MJD=JD$-$2452500). The crosses within the maps indicate the ``centers of light'' of the stars within the other images. Whereas the positions fit the expectations accurately ($\leq 1$ mas), the intensity ratio between the two stars was not always obtained properly due to the poor sampling (the coverage is shown below each map). The contours show $10\%$ intervals scaled to the peak.} \label{fig:mapmotion}
\end{figure*}

\begin{figure*}[hp]
  \epsscale{2.0}
  \includegraphics[width=230pt, height=230pt]{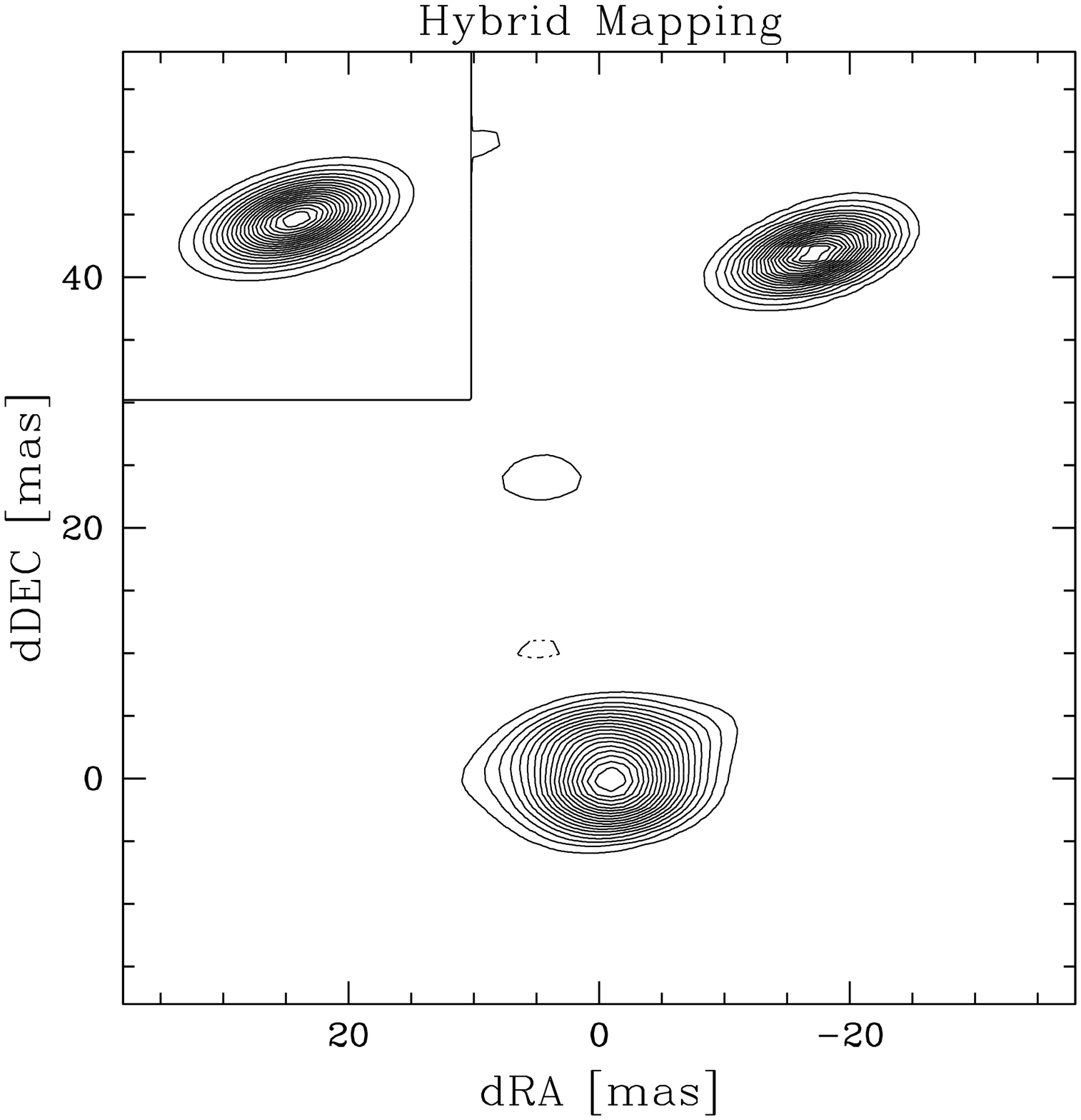}
  \hspace{1mm}
  \includegraphics[width=230pt, height=230pt]{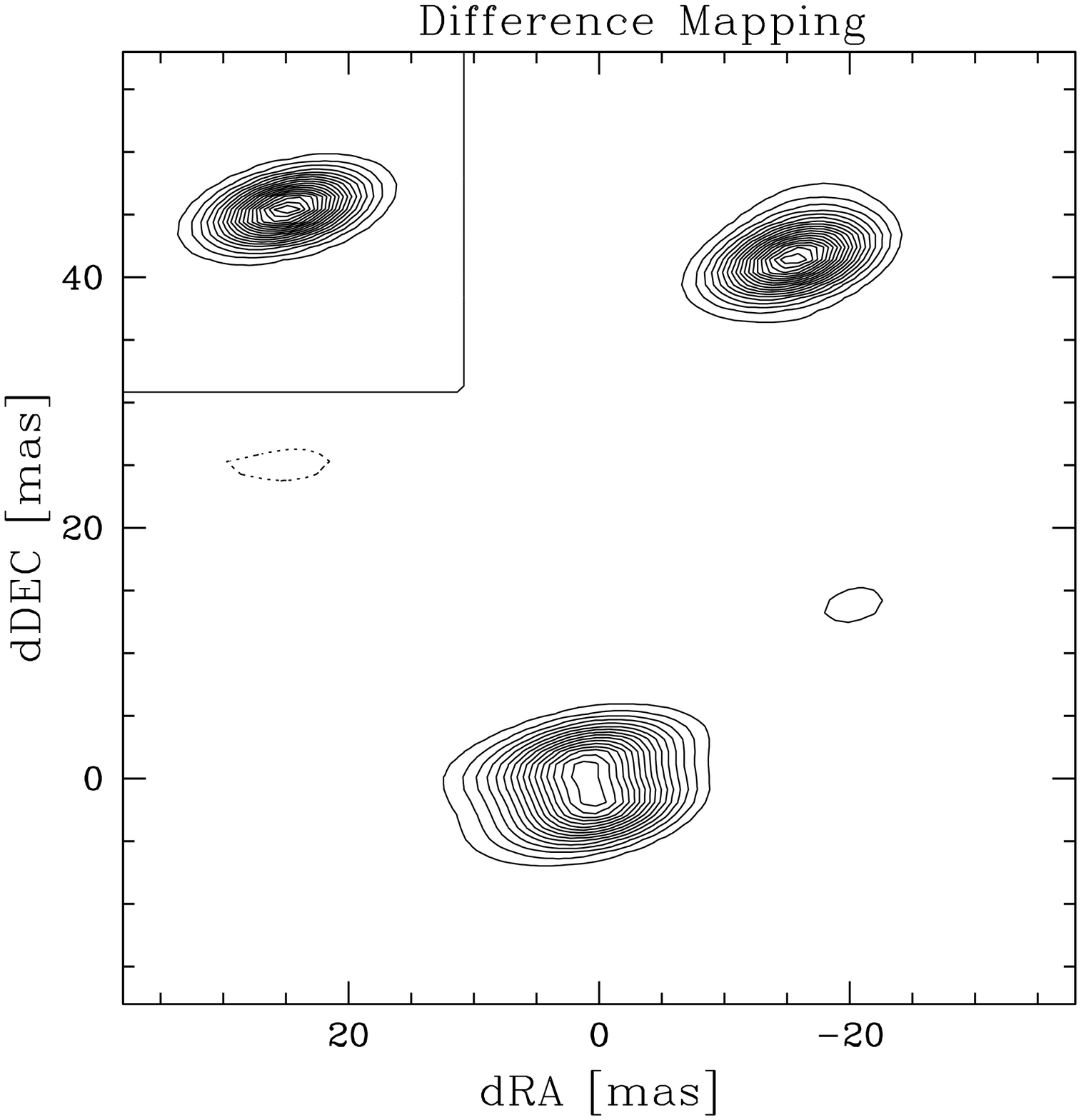}\\
  \vspace{1mm}
  \includegraphics[width=230pt, height=230pt]{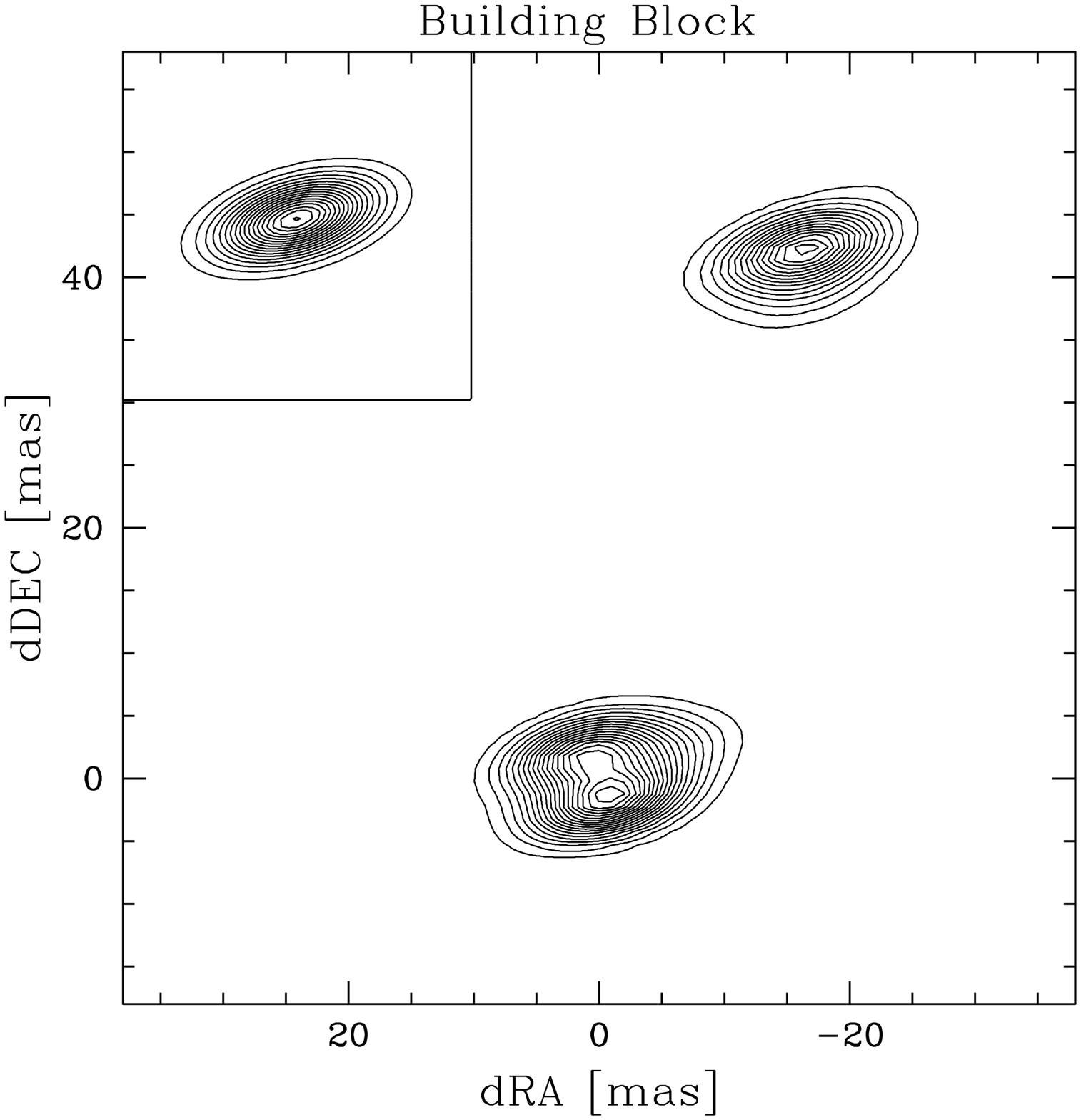}
  \hspace{1mm}
  \includegraphics[width=230pt, height=230pt]{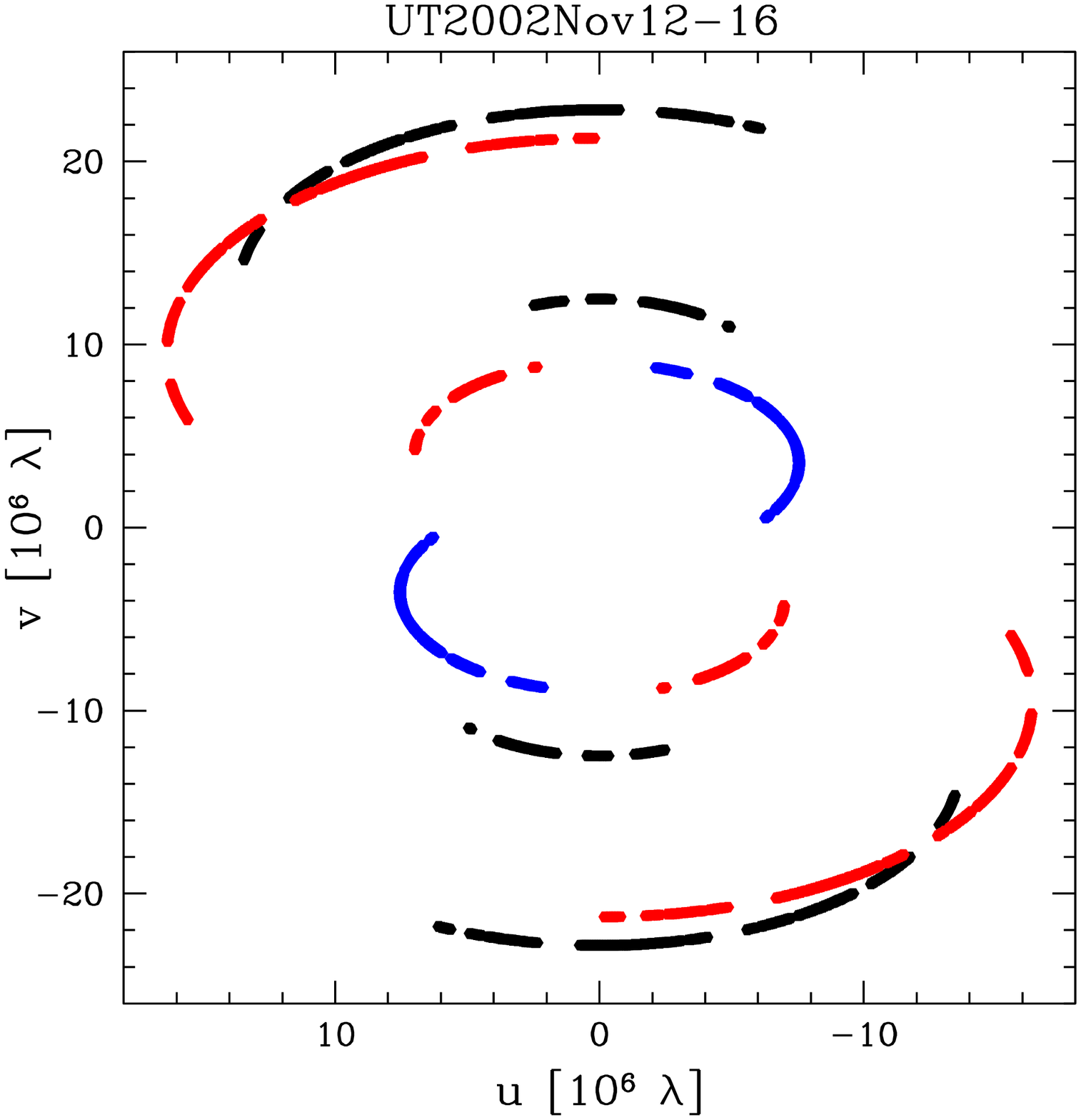}
  \caption{Aperture synthesis maps of Capella generated with data from
  2002\,Nov\,12-16. Maps generated with the CHM (upper left), DFM (upper
  right), and BBM (lower left) algorithm are shown together with the $u,
  v$-plane coverage (lower right). Each map shows the surface brightness
  plotted with $5\%$ interval contours (scaled to the peak intensity). To
  compensate for the motion of the stars over the observed interval ($\approx
  14\degr$), the $u, v$-plane was moved synchronously to the orbital
  motion. The maps are oriented north up, east left. The restoring beam ($5.4
  \times 2.6$ mas) is shown in the upper-left corner of each map as inset.}
  \label{fig:maps}
\end{figure*}

\subsection{Imaging Algorithms used}

In the exploratory spirit of the present Capella imaging work, we have
investigated three different imaging algorithms, as explained here.

The radio astronomy community has developed various algorithms to reconstruct 
phase information from CP, most notably conventional hybrid
mapping \citep[CHM, e.g.][]{cor81} and difference mapping (DFM)
by \cite{pea94}. In both algorithms, the
individual phases are recovered by self-calibration \citep{pea84}. 
The necessary deconvolutions can be performed with the classical CLEAN
algorithm \citep[e.g.][]{hog74}.

We have also tested the ``building-block mapping'' (BBM) algorithm
presented by~\cite{hof93}, which was originally developed to reconstruct
images from speckle-bispectrum data. Here, the image is build up
by adding iteratively point-like components (building blocks) to an image in
order to minimize the least square distance between the measured bispectrum
and the bispectrum of the reconstructed image. Since the computation of this
distance function for each potential position of the next building block (the
whole image space) is computationally too expensive, a linear approximation of
the least square distance is applied. The main assumption of this
approximation is that the added component changes the bispectrum of the
reconstructed image only insignificantly. For details of this method,
see \cite{hof93}.

In order to allow a comparison of the results obtained with the various
methods, we implemented all algorithms within one software 
framework.  This allows us to start all image reconstruction methods
with precisely the same initial conditions.

\subsection{Mapping the orbital motion} \label{cha:mapsubset}

To demonstrate that the data and mapping techniques were sensitive to
the movement of the stars during the observation period, we subdivided our
data into three subsets (2002\,Nov\,12-13, 2002\,Nov\,14, 2002\,Nov\,15-16),
each one still containing enough data for mapping. Since we observed during
the first two epochs with the telescope configuration including the longest
baseline (and hereby missing lower frequencies), the convergence of the BBM
maps had to be supported by limiting the reconstruction area. Within the maps
generated (Fig.~\ref{fig:mapmotion}) it is perceivable that the orbital motion
can clearly be traced accurately. Also the intensity of the stars (from left
to right: $I_{Aa}/I_{Ab} \approx$ 1.6, 1.7, 1.6) can be estimated reliably.

\subsection{Compensating the orbital motion} \label{cha:rotcomp}

The key goal of our project was to produce aperture synthesis images of
Capella in order to demonstrate the capabilities of IOTA/IONIC3. However,
observations were obtained over five nights, during which time the relative
positions of the binary pair changed by many milliarcseconds. In order to make
a single map with data from all nights, and thereby improve the $u, v$ coverage
available for the map, we applied a coodinate transformation to compensate for
the orbital motion of the binary components over the observed time interval.
Thus, the fringe spacing is rescaled to account for the changing separation of
the stars, and the orientation of the fringes on the sky is reoriented to
account for the changing position angle of the binary pair. Each $u, v$ point
was transformed in this manner to a new coordinate system corresponding to the
geometry of the stars at the time of the middle of our observation run. 
This way, all observations may be combined to produce a single map with
improved $u, v$ coverage. 

Of course, the above procedure is only strictly justified for special
circumstances. In a binary system, the stars would have to rotate
synchronously and any brightness structure on each star would have to be
stationary during the time of observation. In the Capella system, we note
that this procedure might be justified for the cooler Aa component, which has
been found to rotate synchronously to the orbital motion~\citep{str01}.
The infrared-fainter Ab component rotates asynchronously, hence it would not
be possible to map structure onto this more active star.

The resulting CHM, DFM and BBM maps are presented in Fig.~\ref{fig:maps}.
For the given, sufficient simple brightness source distribution, all
algorithms converged without any concrete trial model assumptions or other
specific pre-suppositions.  As an initial phase estimate we set two of the
three phases equal to zero, thus the third phase was given by the closure
phase  relation. The intensity ratios $I_{Aa}/I_{Ab}$ measured within the maps
are 1.9 (CHM), 1.6 (DFM), and 1.8 (BBM).

A quantitative comparison between the BBM map in Fig.~\ref{fig:maps} and the
maps constructed from the data subsets as described in
chapter~\ref{cha:mapsubset} reveals that the rotation compensating coordinate
transformation clearly reduced the noise content within the maps and also
improved the convergence behaviour.

Since the stellar surfaces within the maps are partially resolved, one can
measure the stellar diameters. This provides values of
$\Theta_{Aa}^{\textnormal{map}} = 8.3 \pm 1.6$ and 
$\Theta_{Ab}^{\textnormal{map}} = 6.8 \pm 1.2$ (measured within the
unconvolved BBM map), which is in good agreement with our model fitting
results.

\section{Conclusions}

By observing the well-studied Capella system we were able to demonstrate the
overall performance and imaging capabilities of the IOTA/IONIC3 instrument. 
IOTA was able to successfully measure and calibrate visibility amplitudes and
closure phases, even with a temporary but significant instrumental problem
that affected the raw data adversely during this run. The visibility
amplitudes and closure phases were then used with three algorithms to
reconstruct reliable images of this well known binary system.
Thus, the primary goal of this experiment has been achieved, and the IOTA
facility has been shown to be a powerful tool for high-resolution infrared
imaging.

\textit{Acknowledgement.} S.K. was supported by NSF grant AST-0138303 and the
International Max Planck Research School (IMPRS) for Radio and Infrared
Astronomy at the University of Bonn.

\end{document}